\documentclass[prd,nofootinbib,preprint]{revtex4}

\usepackage{amsmath}
\usepackage[dvips]{graphicx}

\newcommand{\capdef}{}
\newcommand{\mycaption}[2][\capdef]{\renewcommand{\capdef}{#2}%
       \caption[#1]{{\footnotesize #2}}}
\makeatletter
\renewcommand{\fnum@table}{\textbf{\tablename~\thetable}}
\renewcommand{\fnum@figure}{\textbf{\figurename~\thefigure}}
\makeatother

\newcommand{\Dmq}{\Delta m^2}
\newcommand{\eVq}{\ensuremath{\text{eV}^2}}
\newcommand{\ON}{{\mathcal O}_N}

\begin{document}

%%%%%%%%%%%%%%%%%%
%%  TITLE PAGE  %%
%%%%%%%%%%%%%%%%%%

\vspace*{10mm}

\title{LSND versus MiniBooNE: Sterile neutrinos with energy dependent
masses and mixing?}

\author{Thomas Schwetz} 
\email{schwetz_at_cern.ch}
\affiliation{Physics Department, Theory Division, CERN, CH--1211
  Geneva 23, Switzerland}

\vspace*{.5cm}

\begin{abstract}
\vspace*{.5cm} Standard active--sterile neutrino oscillations do not
  provide a satisfactory description of the LSND evidence for neutrino
  oscillations together with the constraints from MiniBooNE and other
  null-result short-baseline oscillation experiments. However, if the
  mass or the mixing of the sterile neutrino depends in an exotic way
  on its energy all data become consistent. I explore the
  phenomenological consequences of the assumption that either the
  mass or the mixing scales with the neutrino energy as $1/E_\nu^r$ ($r
  > 0$). Since the neutrino energy in LSND is about 40~MeV, whereas
  MiniBooNE operates at around 1~GeV, oscillations get suppressed in
  MiniBooNE and the two results become fully compatible for $r\gtrsim
  0.2$. Furthermore, also the global fit of all relevant data improves
  significantly by exploring the different energy regimes of the
  various experiments. The best fit $\chi^2$ decreases by 12.7 (14.1)
  units with respect to standard sterile neutrino oscillations if the
  mass (mixing) scales with energy.
\end{abstract}

\preprint{CERN-PH-TH/2007-195}

\maketitle

%%%%%%%%%%%%%%%%%
%%  SECTION 1  %%
%%%%%%%%%%%%%%%%%

\section{Introduction}

Reconciling the LSND evidence~\cite{Aguilar:2001ty} for $\bar\nu_\mu
\to \bar\nu_e$ oscillations with the global neutrino data reporting
evidence~\cite{sk-atm, lbl, sno, kamland} and bounds~\cite{karmen,
Astier:2003gs, Dydak:1983zq, Declais:1994su, Apollonio:2002gd,
Boehm:2001ik} on oscillations remains a long-standing problem for
neutrino phenomenology. Recently the MiniBooNE 
experiment~\cite{MB} added more information to this problem. This
experiment searches for $\nu_\mu\to\nu_e$ appearance with a very
similar $L/E_\nu$ range as LSND. No evidence for oscillations is found
and the results are inconsistent with a two-neutrino oscillation
interpretation of LSND at 98\%~CL~\cite{MB}.
The standard ``solution'' to the LSND problem is to introduce one or
more sterile neutrinos at the eV scale~\cite{sterile}. However, it
turns out that such sterile neutrino schemes do not provide a
satisfactory description of all data in terms of neutrino
oscillations, see Ref.~\cite{Maltoni:2007zf} for a recent analysis
including MiniBooNE data; pre-MiniBooNE analyses can be found, e.g.,
in Refs.~\cite{early, strumia, Maltoni:2002xd, Sorel:2003hf,
Maltoni:2004ei}.
Apart from sterile neutrino oscillations, various more exotic
explanations of the LSND signal have been proposed, for example,
neutrino decay~\cite{Ma:1999im, Palomares-Ruiz:2005vf}, CPT
violation~\cite{cpt, strumia}, violation of Lorentz
symmetry~\cite{lorentz}, CPT-violating quantum
decoherence~\cite{Barenboim:2004wu}, mass-varying
neutrinos~\cite{MaVaN}, or shortcuts of sterile neutrinos in extra
dimensions~\cite{Pas:2005rb}. See Refs.~\cite{Strumia:2006db,
GonzalezGarcia:2007ib} for two recent reviews.

In view of the difficulties to describe all data with ``standard''
sterile neutrino oscillations I assume in this note that the sterile
neutrino is a more exotic particle than just a neutrino without weak
interactions. Being a singlet under the Standard Model gauge group it
seems possible that the sterile neutrino is a messenger from a hidden
sector with some weired properties. I will assume in the following
that the mass or the mixing of the fourth neutrino depends in a
non-standard way on the neutrino energy. The motivation is that the
neutrino energy in LSND is around 40~MeV, whereas MiniBooNE and the
CDHS disappearance experiment operate around 1~GeV, and hence changing
the energy dependence of oscillations will have some impact on the
fit.

To be specific, I am going to assume that either the mass $m_4$ or the
mixing $U_{\alpha 4}$ ($\alpha = e,\mu$) of the fourth neutrino state
depends on the neutrino energy $E_\nu$ like
\begin{equation}\label{eq:scaling}
\begin{aligned}
  \text{MED:}\qquad\qquad  m_4^2(E_\nu) &= 
  \tilde{m}_4^2 \left( \frac{E_\mathrm{ref}}{E_\nu} \right)^r \,,\\ 
  \text{EDM:}\qquad\quad  |U_{\alpha4}|^2(E_\nu) &=
  |\tilde{U}_{\alpha4}|^2 \left( \frac{E_\mathrm{ref}}{E_\nu} \right)^r \,,
\end{aligned}
\end{equation}
where $\tilde{m}_4$ and $\tilde{U}_{\alpha4}$ are the mass and the
mixing at a reference energy $E_\mathrm{ref}$, and for the exponent
$r$ I will consider values in the interval $0\le r \le 1$.  An energy
dependent mass as in Eq.~(\ref{eq:scaling}) will modify the energy
dependence of oscillations from the standard $1/E_\nu$ dependence to
$1/E_\nu^{1+r}$. Therefore, I refer to this effect as modified energy
dependence (MED) oscillations, whereas the second case is denoted by
energy dependent mixing (EDM). Most likely rather exotic physics will
be necessary to obtain such a behaviour, some speculative remarks on
possible origins are given in Sec.~\ref{sec:models}. Here I make the
assumptions of Eq.~(\ref{eq:scaling}) without specifying any
underlying model, instead I will explore the phenomenological
consequences of MED and EDM oscillations for short-baseline neutrino
data. I will show that in both cases ({\it i}) LSND and MiniBooNE
become compatible and ({\it ii}) the global fit improves
significantly.

It is important to note that Eq.~(\ref{eq:scaling}) involves only the
new fourth mass state, whereas masses and mixing of the three active
Standard Model neutrinos are assumed to be constant. Therefore the
successful and very robust description of solar, atmospheric, and
long-baseline reactor and accelerator experiments~\cite{sk-atm, lbl,
sno, kamland} by three-flavour active neutrino oscillations is not
altered significantly, since the mixing of the forth mass state with
active neutrinos is small.

The outline of this paper is as follows.  In Sec.~\ref{sec:framework}
I discuss in some detail the framework of MED and EDM oscillations and
give a qualitative discussion of the expected behaviour of the
combined analysis of the relevant short-baseline oscillation data. In
Sec.~\ref{sec:fit} the results of the numerical analysis are
presented.  The global fit includes the appearance experiments LSND,
MiniBooNE, KARMEN and NOMAD, as well as various disappearance
experiments. In Sec.~\ref{sec:others} I comment briefly on
phenomenological consequences of MED/EDM oscillations in future
oscillation experiments, astrophysics and
cosmology. Sec.~\ref{sec:models} presents some speculative thoughts on
models leading to energy dependent masses and mixing for sterile
neutrinos, and I summarize in Sec.~\ref{sec:summary}.

%%%%%%%%%%%%%%%%%%%%%%%%%%%%%%%%%%%%%%%%%%%%%%%%%%%%%%
\section{The MED and EDM oscillation frameworks}
%%%%%%%%%%%%%%%%%%%%%%%%%%%%%%%%%%%%%%%%%%%%%%%%%%%%%%
\label{sec:framework}

Before discussing qualitatively the phenomenological consequences of
MED and EDM oscillations according to Eq.~(\ref{eq:scaling}) let me
briefly remind the reader about the description of short-baseline
(SBL) neutrino oscillation data in the case of standard four-neutrino
oscillations. In the so-called (3+1) schemes there is a hierarchy
among the mass-squared differences:
\begin{equation}\label{eq:hierarchy}
\Dmq_{21} \ll |\Dmq_{31}| \ll \Delta m^2_{41}\,.
\end{equation}
Under the assumption that $\Dmq_{21}$ and $\Dmq_{31}$ can be
neglected SBL oscillations are described by two-flavour oscillation
probabilities with an effective mixing angle depending on the
elements of the lepton mixing matrix $|U_{e4}|^2$ and $|U_{\mu 4}|^2$.
For $\nu_\mu\to\nu_e$ appearance experiments the effective mixing
angle is given by
\begin{equation}\label{eq:app}
  \sin^2 2\theta_{\mu e} = 4 |U_{e4}|^2 |U_{\mu 4}|^2 \,,
\end{equation}
whereas for a $\nu_\alpha$ disappearance experiment we have
\begin{equation}\label{eq:dis}
  \sin^2 2\theta_{\alpha\alpha} = 4 |U_{\alpha4}|^2 (1 - |U_{\alpha
  4}|)^2 \,,
\end{equation}
see, e.g., Ref.~\cite{Bilenky:1996rw}. The fact that the amplitude
responsible for the LSND appearance is a product of $|U_{e4}|^2$ and
$|U_{\mu 4}|^2$, whereas $\nu_e$ and $\nu_\mu$ disappearance
experiments constrain these elements separately leads to the
well-known tension between LSND and disappearance experiments in the
(3+1) oscillation schemes, see e.g., Refs.~\cite{Bilenky:1996rw,
Okada:1996kw, Barger:1998bn, Bilenky:1999ny, Peres:2000ic,
Grimus:2001mn, cornering}.

Assuming now an energy dependent mass for $\nu_4$ as in
Eq.~(\ref{eq:scaling}), it turns out that for the range of parameters
and energies relevant for our discussion it is always possible to take
$\nu_4$ much heavier than the three standard neutrinos, $m_{1,2,3} \ll
m_4$, such that the usual SBL approximation
Eq.~(\ref{eq:hierarchy}) remains always valid, and the energy scaling
of Eq.~(\ref{eq:scaling}) applies also for the mass-squared difference
$\Delta m^2_{41}(E_\nu) \equiv m^2_4(E_\nu) - m^2_1 \approx
m^2_4(E_\nu)$. Then the relevant oscillation phase $\phi_\mathrm{osc}$
gets a different energy dependence than in the standard case:
\begin{equation}\label{eq:osc_phase}
  \phi_\mathrm{osc} = 
  \frac{\Delta m_{41}^2 L}{4 E_\nu} \approx
  \frac{\Delta \tilde{m}^2_{41} L}{4 E_\nu} 
    \left(\frac{E_\mathrm{ref}}{E_\nu}\right)^r \,.
\end{equation}
Hence the standard $1/E_\nu$ dependence gets altered to
$1/E_\nu^{1+r}$. The most relevant consequence of the MED with $r >
0$ follows from Eq.~(\ref{eq:osc_phase}): An experiment is sensitive
to oscillations if $\phi_\mathrm{osc} \simeq \pi/2$, or
\begin{equation}\label{eq:LoE}
    \frac{L}{E_\nu} \simeq \frac{2\pi}{\Delta\tilde{m}^2_{41}}  
    \left(\frac{E_\nu}{E_\mathrm{ref}}\right)^r \,.
\end{equation}
Hence, for experiments with $E_\nu > E_\mathrm{ref}$ the allowed
region will be shifted to larger values of $\Delta\tilde{m}^2_{41}$ as
compared to the standard oscillation case ($r = 0$), whereas the
allowed region for experiments with $E_\nu < E_\mathrm{ref}$ will be
shifted to smaller $\Delta\tilde{m}^2_{41}$, in order to compensate
for the factor $(E_\nu / E_\mathrm{ref})^r$ in Eq.~(\ref{eq:LoE}).

Using instead of the MED now the energy dependence of the mixing
matrix elements from Eq.~(\ref{eq:scaling}), one obtains for the SBL
appearance and disappearance amplitudes given in Eqs.~(\ref{eq:app})
and (\ref{eq:dis}):
\begin{equation}
  \begin{aligned}
    \sin^22\theta_{\mu e} &= 4 |U_{e4}|^2 |U_{\mu 4}|^2 
    \propto \left( \frac{E_\mathrm{ref}}{E_\nu} \right)^{2r} \,,\\
    \sin^22\theta_{\alpha\alpha} &= 4 |U_{\alpha4}|^2 (1 - |U_{\alpha 4}|^2)
    \approx 4 |U_{\alpha4}|^2 
    \propto \left( \frac{E_\mathrm{ref}}{E_\nu} \right)^r \,.
  \end{aligned}
\end{equation}
This introduces only a mild distortion of the oscillation pattern from
the standard oscillatory behaviour with $1/E_\nu$. The main effect of
EDM is that the sensitivity of experiments with $E_\nu >
E_\mathrm{ref}$ to the effective mixing angle gets weaker.  Let us
note that the EDM scaling of Eq.~(\ref{eq:scaling}) cannot hold for
arbitrarily low energies, simply because of unitarity of $U$. The low
energy limit of the EDM scaling should find an explanation in some
theory for this effect. Here I assume that for the parameter range
relevant for the SBL analysis the power-law scaling of
Eq.~(\ref{eq:scaling}) remains valid. 

Note that in both cases, MED and EDM, for fixed $r$ the choice of the
reference energy $E_\mathrm{ref}$ is arbitrary. From
Eq.~(\ref{eq:osc_phase}) it is clear that choosing a different
reference energy $E_\mathrm{ref}$ leads just to a rescaling of
$\tilde{m}_4$ such that the combination $\Delta \tilde{m}^2_{41}
E_\mathrm{ref}^r$ remains constant. Similar, changing $E_\mathrm{ref}$
in case of the EDM leads just to a rescaling of the
$|\tilde{U}_{\alpha 4}|^2$.  Hence, the (3+1) MED and EDM models have
one phenomenological parameter in addition to the (3+1) standard
oscillation model: 
\begin{equation}\begin{aligned}
  \text{MED:}\qquad&
  |U_{e4}|^2,\, |U_{\mu 4}|^2,\, \Delta \tilde{m}^2_{41},\, r\,,\\
  \text{EDM:}\qquad&
  |\tilde{U}_{e4}|^2,\, |\tilde{U}_{\mu 4}|^2,\, \Delta m^2_{41},\, r\,.
\end{aligned}\end{equation}

The main effects of MED and EDM oscillations can be summarized in the
following way: Consider the allowed regions of the various experiments
for standard oscillations in the plane of $\sin^22\theta_{\mu e}$ and
$\Dmq_{41}$. Introducing now MED (EDM) leads to a relative shift of
the regions of experiments at different energies along the
$\Delta\tilde{m}^2_{41}$ ($\sin^22\tilde{\theta}_{\mu e}$) axis.
The relevant SBL experiments are listed in Tab.~\ref{tab:experiments},
ordered according to their mean neutrino energy. For convenience I
will choose $E_\mathrm{ref} = 40$~MeV, corresponding roughly to the
mean neutrino energy in LSND. Since in MiniBooNE the neutrino energy
is higher, for MED (EDM) oscillations the sensitivity is shifted to
larger values of $\Delta\tilde{m}^2_{41}$ ($\sin^22\tilde{\theta}_{\mu
e}$) for $r > 0$. In the MED framework MiniBooNE operates actually at
a too small value of $L/E_\nu$ in order to test LSND. As I will show
in the following, in both cases the two experiments become compatible
for $r > 0$. Furthermore, it turns out that also the global fit
including all the experiments listed in the table improves
significantly due to the different energy regimes.

\begin{table}[t] \centering
    \begin{tabular}{l@{\quad}c@{\quad}c@{\quad}c@{\quad}c}
	\hline\hline
        Experiment & Ref. & Channel & Data & $\langle E_\nu \rangle$
        \\
	\hline
	Bugey     & \cite{Declais:1994su}  &$\bar\nu_e\to\bar\nu_e$ & 60 &   4 MeV \\
	Chooz     & \cite{Apollonio:2002gd}&$\bar\nu_e\to\bar\nu_e$ &  1 &   4 MeV \\
	Palo Verde& \cite{Boehm:2001ik}    &$\bar\nu_e\to\bar\nu_e$ &  1 &   4 MeV \\
	\hline
	LSND      & \cite{Aguilar:2001ty}  &$\bar\nu_\mu\to\bar\nu_e$& 11 &  40 MeV \\
        KARMEN    & \cite{karmen}          &$\bar\nu_\mu\to\bar\nu_e$&  9 &  40 MeV \\
	\hline
        MiniBooNE & \cite{MB}              &$\nu_\mu\to\nu_e$       &  8 & 700 MeV \\
        CDHS      & \cite{Dydak:1983zq}    &$\nu_\mu\to\nu_\mu$     & 15 &   1 GeV \\
        NOMAD     & \cite{Astier:2003gs}   &$\nu_\mu\to\nu_e$       &  1 &  50 GeV \\
	\hline\hline
    \end{tabular}
    \mycaption{\label{tab:experiments} Experiments used in the SBL
     analysis. The oscillation channel, and the number of data points
     used in the fit (``Data'') are given. The last column shows the
     approximate mean neutrino energy for each experiment.}
\end{table}

%%%%%%%%%%%%%%%%%%%%%%%%%%%%%%%%%%%%%%%%%%%%%%%%%%%%%%%%%%%%%%%%%%
\section{Global analysis of SBL data for MED and EDM oscillations}
%%%%%%%%%%%%%%%%%%%%%%%%%%%%%%%%%%%%%%%%%%%%%%%%%%%%%%%%%%%%%%%%%%
\label{sec:fit}

\subsection{Description of the data used in the fit}

Before presenting the results of the analysis let us briefly discuss
the data used in the fit, as summarized in Tab.~\ref{tab:experiments}.
For the re-analysis of LSND I fit the observed transition
probability (total rate) plus 11 data points of the $L/E$ spectrum
with free normalisation, both derived from the decay-at-rest
data~\cite{Aguilar:2001ty}. For KARMEN the data observed in 9 bins of
prompt energy as well as the expected background~\cite{karmen} is used
in the fit.  Further details of the LSND and KARMEN analyses are given
in Ref.~\cite{Palomares-Ruiz:2005vf}. For NOMAD I fit the total rate
using the information provided in Ref.~\cite{Astier:2003gs}.

The MiniBooNE analysis is based on the information provided at the
web-page \cite{MB-data}, derived from the actual Monte Carlo simulation
performed by the collaboration. Using these data the ``official''
MiniBooNE analysis~\cite{MB} can be reproduced with very good accuracy. The
averaging of the transition probability is performed with the proper
reconstruction efficiencies, and detailed information on error
correlations and backgrounds is available for the $\chi^2$ analysis.
MiniBooNE data are consistent with zero (no excess) above 475~MeV,
whereas below this energy a $3.6\sigma$ excess of $96 \pm 17 \pm 20$
events is observed. Whether this excess comes indeed from
$\nu_\mu\to\nu_e$ transitions or has some other origin is under
investigation.  As discussed in Ref.~\cite{MB}, standard two-neutrino
oscillations cannot account for the event excess at low
energies. Following the strategy of the MiniBooNE collaboration the
analysis is restricted to the energy range from 475~MeV to 3~GeV. In
Sec.~\ref{sec:MB300} I will comment on the possibility to obtain the
low energy event excess in case of MED oscillations.

I include the disappearance experiments Bugey~\cite{Declais:1994su},
Chooz~\cite{Apollonio:2002gd}, and Palo Verde~\cite{Boehm:2001ik}
(reactor $\bar\nu_e$ disappearance), as well as the
CDHS~\cite{Dydak:1983zq} $\nu_\mu$ disappearance experiment, see
Ref.~\cite{Grimus:2001mn} for technical details.
In addition to the data listed in Tab.~\ref{tab:experiments}
atmospheric neutrino data give an important constraint on the mixing
of $\nu_\mu$ with the heavy mass state, i.e., on $|U_{\mu
4}|^2$~\cite{Bilenky:1999ny, cornering}. I use the updated analysis
described in detail in Ref.~\cite{Maltoni:2007zf}. The basic
assumption in this analysis is that $\Dmq_{41}$ is ``infinite'' for
atmospheric neutrinos according to Eq.~(\ref{eq:hierarchy}). Since I
assume this to be true also in the scenarios under consideration one
can directly apply the bound on $|U_{\mu 4}|^2$ from the standard
oscillation analysis. In case of EDM one should perform a re-analysis
of atmospheric data taking into account the energy dependence of
$|U_{\mu4}|^2$ for the various data samples spanning five decades in
neutrino energy. Such an analysis is beyond the scope of the present
work and I assume a scaling of $|U_{\mu4}|^2$ corresponding to an
average energy of 1~GeV. Adding one data point for the bound from
atmospheric neutrinos to the data given in Tab.~\ref{tab:experiments}
I obtain $N_\mathrm{tot} = 107$ data points in the global analysis.

%%%%%%%%%%%%%%%%%%%%%%%%%%%%%%%%%%%%%%%%%%%%%%%%%%%%%
\subsection{Results of the global analysis}

\begin{figure}[t] \centering 
    \includegraphics[width=0.9\textwidth]{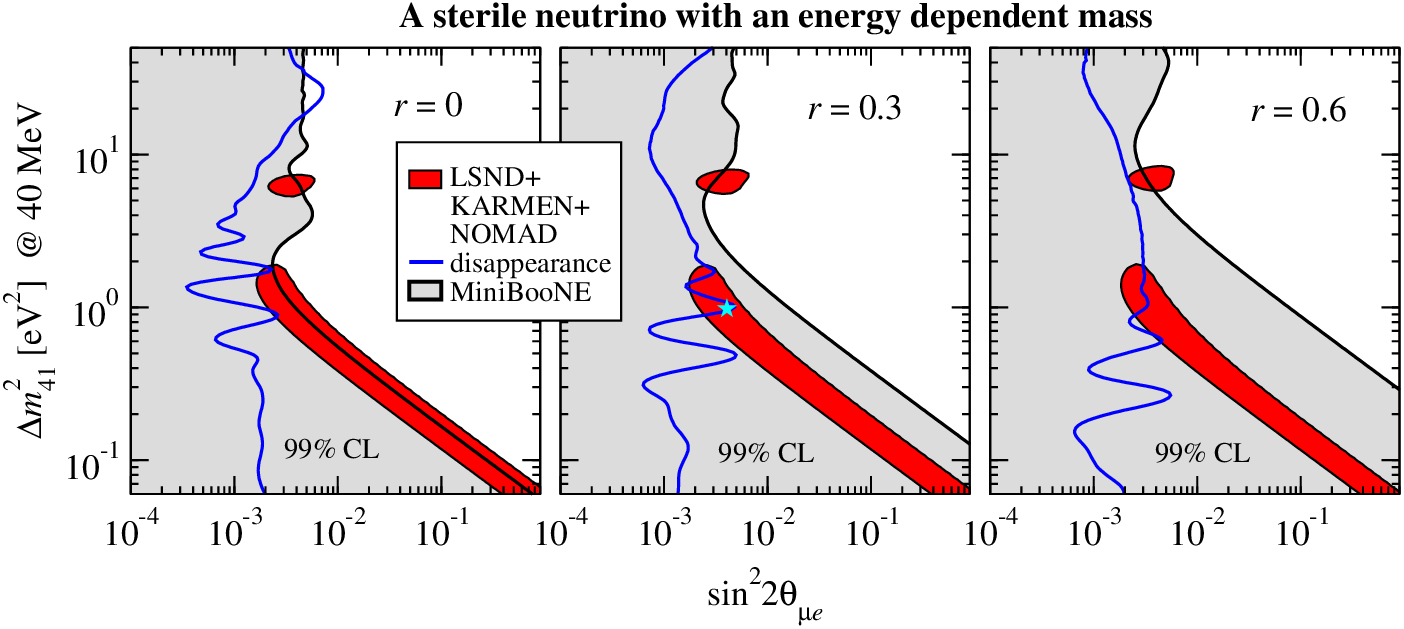}
    \mycaption{\label{fig:MED} (3+1) oscillations with a modified
    energy dependence (MED): Allowed regions at 99\%~CL (2~d.o.f.) for
    LSND+KARMEN+NOMAD, MiniBooNE, and the disappearance experiments
    for standard oscillations $r = 0$ (left) and MED oscillations with
    $r = 0.3$ (middle) and $r = 0.6$ (right). The star in the middle
    panel corresponds to the global best fit point.}
\end{figure}

\begin{figure}[t] \centering 
    \includegraphics[width=0.9\textwidth]{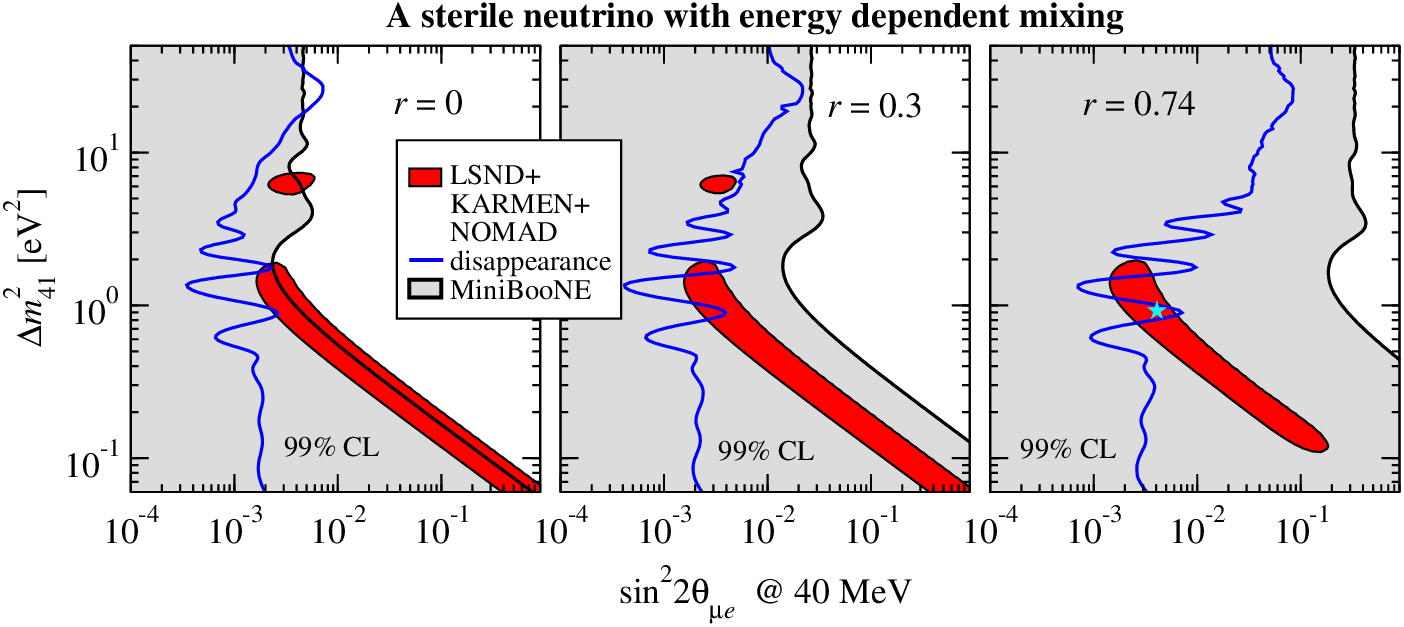}
    \mycaption{\label{fig:EDM} (3+1) oscillations with energy
    dependent mixing (EDM): Allowed regions at 99\%~CL (2~d.o.f.) for
    LSND+KARMEN+NOMAD, MiniBooNE, and the disappearance experiments
    for standard oscillations $r = 0$ (left) and EDM oscillations with
    $r = 0.3$ (middle) and $r = 0.74$ (right). The star in the right
    panel corresponds to the global best fit point.}
\end{figure}

Let us now discuss the results of the numerical analysis within the
frameworks of MED and EDM oscillations. Figs.~\ref{fig:MED} and
\ref{fig:EDM} show the allowed regions in the plane of $\Dmq_{41}$ and
$\sin^22\theta_{\mu e}$ for various data sets for standard
oscillations ($r = 0$) compared to the exotic energy dependence models
for some values of $r > 0$. First, note that the neutrino energy in
LSND and KARMEN is the same, and therefore the consistency of LSND and
KARMEN~\cite{Church:2002tc} is not affected by introducing a non-zero
$r$. Furthermore, since we choose a reference energy $E_\mathrm{ref}$
close to the mean energy in these experiments the allowed region from
LSND+KARMEN+NOMAD does not change by increasing $r$.\footnote{NOMAD,
with an energy of about 50~GeV, contributes very little to these
regions.}
Second, in agreement with the argument given in
Sec.~\ref{sec:framework} one observes from the figures that the bound
from MiniBooNE moves to higher values of $\Dmq_{41}$ for MED and to
higher values of $\sin^22\theta_{\mu e}$ for EDM if $r$ increases, as
a consequence of the higher neutrino energy in MiniBooNE. I find that
in both cases for $r \gtrsim 0.2$ LSND and MiniBooNE are fully
consistent.
Third, Figs.~\ref{fig:MED} and \ref{fig:EDM} show the bound on
$\sin^22\theta_{\mu e}$ from the disappearance experiments Bugey,
Chooz, Palo Verde, CDHS, and atmospheric neutrino data, where for a
given $\sin^22\theta_{\mu e}$ I minimize the $\chi^2$ with respect to
$|U_{e4}|^2$ and $|U_{\mu 4}|^2$ under the constraint
$\sin^22\theta_{\mu e} = 4 |U_{e4}|^2 |U_{\mu 4}|^2$.  As visible in
the left panels, in the standard oscillation case there is severe
tension between these data and the LSND evidence, and at 99\%~CL there
is basically no overlap of the allowed regions. However, the situation
clearly improves for MED and EDM oscillations, and for $r > 0$ the
allowed regions overlap.

For MED oscillations (Fig.~\ref{fig:MED}) this can be understood in
the following way. The pronounced wiggles in the disappearance bound
visible in the left panel around $\Dmq_{41} \sim 1$~eV$^2$ come from
the Bugey reactor experiment.  Since $E_\nu$ for Bugey is smaller than
the reference energy of 40~MeV these features move to lower values of
$\Delta\tilde{m}^2_{41}$ if $r$ is increased. On the other hand, the
constraint on $|U_{\mu 4}|^2$ from CDHS is shifted to higher values of
$\Delta\tilde{m}^2_{41}$ and only the weaker constraint from
atmospheric data remains. Both trends work together in moving the
disappearance bound towards the LSND region, as visible in the middle
panel for $r = 0.3$. If $r$ is further increased the Bugey pattern
moves to even smaller values of $\Delta\tilde{m}^2_{41}$, and the
bound at the relevant region around 1~eV$^2$ is given by the
constraints from Chooz on $|U_{e 4}|^2$ and atmospheric neutrinos on
$|U_{\mu 4}|^2$ (see right panel).

In the case of EDM oscillations (Fig.~\ref{fig:EDM}) there are two
opposite trends. Since for Bugey $E_\nu < E_\mathrm{ref}$ the
constraint on $|\tilde{U}_{e4}|^2$ becomes stronger with increasing
$r$, whereas for CDHS and atmospheric data $E_\nu > E_\mathrm{ref}$
and the bound on $|\tilde{U}_{\mu4}|^2$ gets weaker. The upper limit
on $\sin^22\tilde{\theta}_{\mu e}$ emerges from the product of these two
constraints, see Eq.~(\ref{eq:app}), and therefore, it scales
according to
\begin{equation}\label{eq:scaling-bound}
\left(\frac{E_\mathrm{ref}^2}
{\langle E_\nu \rangle_\mathrm{Bugey} \,
 \langle E_\nu \rangle_\mathrm{CDHS,atm}}
\right)^r
\simeq
0.4^r \,.
\end{equation}
Hence, the net-effect is a shift of the bound to larger values of
$\sin^22\tilde{\theta}_{\mu e}$ and a significant overlap with the
LSND region emerges.

\begin{figure}[t] \centering 
    \includegraphics[width=0.7\textwidth]{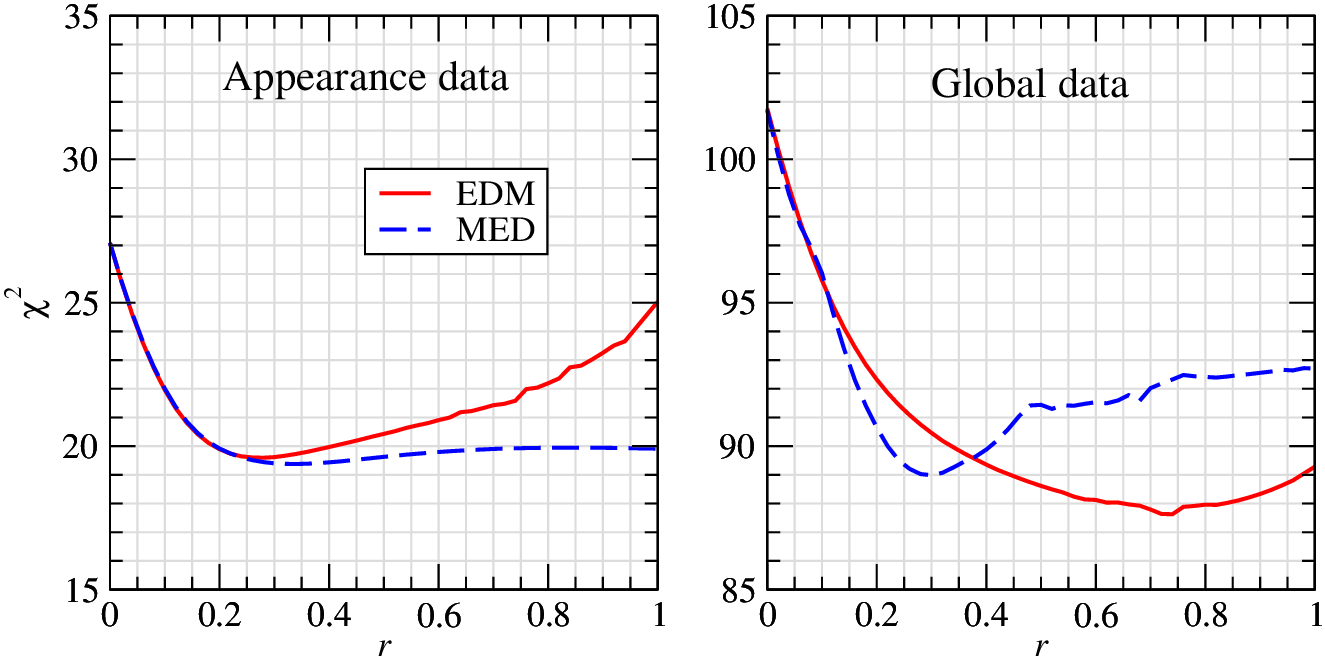}
    \mycaption{\label{fig:chisq} The $\chi^2$ for SBL appearance data
    (left) and global data (right) for MED and EDM oscillations as a
    function of the exponent $r$, minimized with respect to the other
    parameters. The number of data points is 29 for the appearance
    experiments and 107 for the global analysis.}
\end{figure}

Fig.~\ref{fig:chisq} shows the $\chi^2$ for appearance data only (left)
and for the global data (right) as a function of the energy exponent $r$. 
The best fit points have the following $\chi^2$ values:
\begin{equation}\begin{aligned}
  \text{MED:}\qquad&
  \chi^2_\mathrm{app,min}  = 19.4/(29-3)\,\text{d.o.f.} \,,\qquad
  \chi^2_\mathrm{glob,min} = 89.0/(107-4)\,\text{d.o.f.} \,, \\
  \text{EDM:}\qquad&
  \chi^2_\mathrm{app,min}  = 19.6/(29-3)\,\text{d.o.f.} \,,\qquad
  \chi^2_\mathrm{glob,min} = 87.6/(107-4)\,\text{d.o.f.} \,,
\end{aligned}\end{equation}
and occur at the parameter values
\begin{equation}\label{eq:best-fit}
\begin{aligned}
  \text{MED:}\qquad& 
  |U_{e4}| = 0.15,\, |U_{\mu 4}| = 0.21,\, 
  \Delta \tilde{m}^2_{41} = 1.0~\eVq,\, r=0.3 \,,\\ 
  \text{EDM:}\qquad&
  |\tilde{U}_{e4}| = 0.06,\, |\tilde{U}_{\mu 4}| = 0.53,\, 
  \Dmq_{41} = 0.92~\eVq,\, r=0.74 \,.
\end{aligned}\end{equation}
As mentioned above, the parameters with tilde correspond always to a
reference energy of 40~MeV. In agreement with the discussion above one
observes that for EDM $|\tilde{U}_{e4}|$ is rather small to respect
the stronger bound from Bugey, whereas $|\tilde{U}_{\mu 4}|$ gets
relatively large due to the relaxed bound from CDHS and atmospheric
data. Although the best fit point for EDM occurs at the relatively
large value of $r = 0.74$, one can see from Fig.~\ref{fig:chisq} that
fits of comparable quality are obtained already for $r \gtrsim 0.4$.

The MED and EDM fits improves significantly with respect to the standard
oscillation case $r = 0$:
\begin{equation}\begin{aligned}
  \text{MED:}\qquad&
  \Delta\chi^2_\mathrm{app}(r=0) = 7.7 \,,\qquad
  \Delta\chi^2_\mathrm{glob}(r=0) = 12.7 \,,\\
  \text{EDM:}\qquad&
  \Delta\chi^2_\mathrm{app}(r=0) = 7.5 \,,\qquad
  \Delta\chi^2_\mathrm{glob}(r=0) = 14.1 \,.\\
\end{aligned}\end{equation}
For comparison, the extension of the standard (3+1) oscillation scheme
to a (3+2) scheme by the addition of a second sterile neutrino leads
to an improvement of $\chi^2_\mathrm{min\,(3+1)} -
\chi^2_\mathrm{min\,(3+2)} = 6.1$~\cite{Maltoni:2007zf}. Taking into
account that for (3+2) oscillations 4 additional parameters are
introduced in the fit instead of only one as in the cases of MED or
EDM, one concludes that the latter provide a much more significant
improvement of the fit.

The shape of the curves in Fig.~\ref{fig:chisq} can be understood from
the behaviour of the allowed regions shown in Figs.~\ref{fig:MED} and
\ref{fig:EDM}. The best fit for appearance data is reached once the
MiniBooNE exclusion curve is moved out of the LSND region, and no
further improvement can be obtained by further increasing $r$.  In the
case of EDM the fit gets worse again due to the energy distortion
introduced for large $r$ by scaling the mixing with $1/E_\nu^r$. This is
also the reason for the change in the LSND region visible in the right
panel of Fig.~\ref{fig:EDM}.
For MED, the global $\chi^2$ reaches a minimum when the wiggles from
the Bugey bound cover the LSND region around $\Delta\tilde{m}^2_{41}
\simeq 1~\eVq$. If $r$ is further increased these wiggles are moved out
again of the LSND region and the fit gets slightly worse again. For $r
\gtrsim 0.5$ a plateau is reached, since then the disappearance bound
at $\Delta\tilde{m}_{41} \simeq 1~\eVq$ comes mainly from Chooz and
atmospheric data, which are independent of $\Dmq_{41}$ and hence also
independent of $r$.
In the case of EDM, the global $\chi^2$ improves until relatively
large values of $r$ as a consequence of
Eq.~(\ref{eq:scaling-bound}). At some point again the fit gets worse
due to the anomalous energy dependence of the probability.

A powerful tool to evaluate the compatibility of different data sets
is the so-called parameter goodness-of-fit (PG) criterion discussed in
Ref.~\cite{Maltoni:2003cu}. It is based on the $\chi^2$ function
\begin{equation} \label{eq:PG}
    \chi^2_\text{PG} = 
    \chi^2_\text{tot,min} - \sum_i \chi^2_{i,\text{min}} \,,
\end{equation}
where $\chi^2_\text{tot,min}$ is the $\chi^2$ minimum of all data sets
combined and $\chi^2_{i,\text{min}}$ is the minimum of the data set
$i$. This $\chi^2$ function measures the ``price'' one has to pay by
the combination of the data sets compared to fitting them
independently. It should be evaluated for the number of d.o.f.\
corresponding to the number of parameters in common to the data sets,
see Ref.~\cite{Maltoni:2003cu} for a precise definition. 

\begin{table}[t] \centering
    \begin{tabular}{l@{\quad}cc@{\quad}cc@{\quad}cc}
	\hline\hline
        & \multicolumn{2}{c}{Standard oscillations}
	& \multicolumn{2}{c}{MED oscillations}
	& \multicolumn{2}{c}{EDM oscillations}
        \\
	Data sets & $\chi^2_\mathrm{PG}/$d.o.f. & PG 
	          & $\chi^2_\mathrm{PG}/$d.o.f. & PG 
	          & $\chi^2_\mathrm{PG}/$d.o.f. & PG 
	\\
	\hline
        LSND vs NEV 
	& 24.9/2 & $4\times 10^{-6}$ 
        & 14.0/3 & $0.3\%$
        & 11.9/3 & $0.8\%$
	\\
	LSND vs NEV-APP vs DIS
	& 25.3/4 & $4\times 10^{-5}$ 
        & 14.3/6 & $3\%$
        & 12.3/6 & $5\%$
	\\
	LKN vs MiniBooNE vs DIS
	& 20.1/4 & $5\times 10^{-5}$ 
        &  8.9/6 & $18\%$
        &  6.7/6 & $35\%$
	\\
	\hline\hline
    \end{tabular}
    \mycaption{\label{tab:PG} Consistency tests of various data
      sub-sets for standard (3+1) oscillations, MED and EDM
      oscillations. The data sets are SBL data showing no evidence for
      oscillations (NEV), no-evidence appearance data (NEV-APP =
      MiniBooNE + KARMEN + NOMAD), SBL disappearance data (DIS), and
      LSND + KARMEN + NOMAD (LKN). I give $\chi^2_\mathrm{PG}$
      according to Eq.~(\ref{eq:PG}) and the corresponding probability
      (``PG'').}
\end{table}

The results of such a PG analysis are displayed in Tab.~\ref{tab:PG}.
First, the compatibility of LSND and all the remaining no-evidence SBL
data is tested, and the PG is compared within the standard, the MED,
and the EDM oscillation frameworks. The consistency improves
drastically from $4\times 10^{-6}$ to $3\times 10^{-3}$ (MED) or
$8\times 10^{-3}$ (EDM). The probability value in the MED case
corresponds to a tension of slightly less than $3\sigma$.  Second, as
an alternative test I check the compatibility of the three data sets
LSND, no-evidence appearance data, and disappearance data. Similar a
huge improvement of the consistency from $4\times 10^{-5}$ for
standard oscillations to 3\% (MED) or 5\% (EDM) is found. 

Although both of these two tests clearly show an improvement of the
fit with respect to standard oscillations, the low probabilities still
indicate that the global fit is not perfect and some tension remains
in the data. The reason for this is the tension between KARMEN and
LSND, which is the same as in the standard oscillation case, since
these experiments have the same energy. The allowed regions in the
space of oscillation parameters of these two experiments have clearly
an overlap, and a careful combined analysis came to the conclusion
that they are consistent~\cite{Church:2002tc}.  Nevertheless, there
remains a tension between them which is detected by the rather
sensitive PG test. In the last row of Tab.~\ref{tab:PG} I consider the
case when LSND and KARMEN are combined to one single data set (which
includes also NOMAD), and test the consistency of this set against
MiniBooNE and disappearance data. This corresponds to the data sets
shown in Figs.~\ref{fig:MED} and \ref{fig:EDM}, and since LSND and
KARMEN are included in the same data set the tension between them does
not show up in the PG. In this case the PG shows a perfect consistency
of all data with probablities of 18\% and 35\% for MED and EDM,
respectively, whereas the probablitiy of standard oscillations remains
unacceptably low.

\subsection{The low energy excess in MiniBooNE}
\label{sec:MB300}

Before concluding this section I comment briefly on the event excess
observed in MiniBooNE in the energy region below 475~MeV. As discussed
in Ref.~\cite{MB}, standard two-flavour oscillations cannot account
for the sharp rise at low energy. However, since in the MED scenario
the energy dependence of oscillations is modified according to
Eq.~(\ref{eq:osc_phase}) one may expect that an explanation of the
excess becomes possible. Indeed I find that for values of the exponent
$r \gtrsim 1$ the rise of the oscillation probability becomes steep
enough and a perfect fit to the full spectrum including the excess
between 300 and 475~MeV becomes possible.  For such large values of
$r$ a closed allowed region appears in the plane of
$\Delta\tilde{m}^2_{41}$ and $\sin^22\theta_{\mu e}$ for MiniBooNE
data (not only a bound). However, because of the large $r$ value this
allowed region appears at $\Delta\tilde{m}^2_{41}$ values above the
LSND region and the KARMEN bound. Hence, although a modified energy
dependence like considered here allows in principle for an explanation
of the low energy event excess, this solution is not compatible with
LSND and the KARMEN bound, and in the global analysis the excess
cannot be fitted.  For this reason I used in the analysis only the
MiniBooNE data above 475~MeV~\cite{MB}, and rely on an alternative
explanation of the low energy event excess.

%%%%%%%%%%%%%%%%%%%%%%%%%%%%%%%%%%%%%%%%%
\section{Implications for future experiments, astrophysics and cosmology}
%%%%%%%%%%%%%%%%%%%%%%%%%%%%%%%%%%%%%%%%%
\label{sec:others}

In this section I briefely comment on other phenomenological
implications of the MED/EDM schemes.
In general the scenarios considered here are difficult to test at
future neutrino oscillation experiments. No appearance signal is
expected for MiniBooNE anti-neutrino data, which currently are being
accumulated, since the energy dependence is assumed to affect
anti-neutrinos in the same way as neutrinos. In order to test the
LSND signal for a MED with $r = 0.3$ at the given MiniBooNE baseline
one would need to run at an energy of
\begin{equation}
  E_\mathrm{MiniBooNE}^\mathrm{MED} \simeq
  \left(\frac{L_\mathrm{MiniBooNE}}{L_\mathrm{LSND}}\right)^\frac{1}{1+r} 
  E_\mathrm{LSND}
  \approx 
  \left(\frac{540\,\mathrm{m}}{30\,\mathrm{m}}\right)^{0.77} 40\,\mathrm{MeV}
  \approx 360\,\mathrm{MeV}\,, 
\end{equation}
which seems not practicable because of low cross sections and large
backgrounds. A similar signal as in standard (3+1) or (3+2)
oscillations is expected also in the MED scenario for future reactor
experiments, see Ref.~\cite{Bandyopadhyay:2007rj}.  A promising place
to look for effects of MED oscillations could be the 2~km detector
proposed for the T2K experiment~\cite{Itow:2001ee}. With a mean
neutrino energy of 0.7~GeV this detector is slightly too far from the
neutrino source to cover the oscillation maximum in case of standard
oscillations with $\Dmq \sim 1$~eV$^2$. However, with the MED best fit
point from Eq.~(\ref{eq:best-fit}) the oscillation phase according to
Eq.~(\ref{eq:osc_phase}) turns out to be close to $\pi/2$ at $L=2$~km.

A rather model independent test of MED or EDM explanations of LSND
would be an experiment operating at the same energy as LSND, such as
proposed in Ref.~\cite{Garvey:2005pn}. Furthermore, to test the EDM
scenario one would like to perform experiments at energies as small as
possible. In particular, this model predicts relatively large effects
for $\nu_\mu$ disappearance for experiments at energies smaller than
1~GeV, since the eV-scale mass state has a rather large mixing with
$\nu_\mu$ at 40~MeV, see Eq.~(\ref{eq:best-fit}).
Let us note that unitarity requires $|U_{\alpha 4}| \le 1$. Therefore,
the power law energy dependence of EDM cannot hold down to arbitrarily
low energies. At the best fit values given in Eq.~(\ref{eq:best-fit}),
one finds $|U_{e4}| \simeq 1$ at $E_\nu \simeq 0.02$~MeV and $|U_{\mu
4}| \simeq 1$ at $E_\nu \simeq 7$~MeV.\footnote{Note that for the SBL
analysis only $|U_{e4}|$ is needed at few~MeV energies, whereas
$|U_{\mu 4}|$ is evaluated only for $E_\nu \gtrsim 40$~MeV.} In order
to use the EDM framework for very low energies one would have to
specify the energy dependence of the neutrino mass matrix, and obtain
the mixing angles via the diagonalisation, such that unitarity is
always guaranteed. This is especially relevant, for example, to obtain
predictions for neutrino mass experiments from Tritium beta-decay,
which has an end point energy of 18.6~keV.

As in case of standard sterile neutrino mixing, also in the MED/EDM
framework the sterile neutrinos have implications for cosmology and
astrophysics, see Ref.~\cite{Cirelli:2004cz} and references
therein. In general the effects will be very similar to the standard
case with effective masses and mixing evaluated according to the
relevant neutrino energy. For example, in a supernova and in Big Bang
nucleosynethesis (BBN) the neutrino energy is close to (or slightly
below) the reference energy 40~MeV used above, and therefore the
mixing parameters shown in Eq.~(\ref{eq:best-fit}) roughly apply in
these environments. This implies that---as in the standard (3+1)
case---the sterile neutrino will be brought into thermal equilibrium
via oscillations prior to BBN~\cite{Okada:1996kw, DiBari:2003fg}.

Cosmology provides a bound on the sum of the neutrino masses in the
sub-eV range~\cite{cosmo-bounds}, mainly from the power spectrum at
large scales combined with precise data on the cosmic microwave
background. In general this bound implies a challenge for sterile
neutrino schemes relevant for LSND. The conflict becomes particularly
severe for the MED framework, since here the neutrino mass increases
with decreasing neutrino energy, which implies large masses for
cosmological relevant neutrinos. Let us note, however, that depending
on the particular model realisation of MED one can expect that the
power law scaling of Eq.~(\ref{eq:scaling}) does not continue down to
arbitrarily low energies. Here I assume only that it holds in the
energy interval relevant for SBL experiments, i.e., above about 1~MeV,
and it might be altered at lower energies.

%%%%%%%%%%%%%%%%%%%%%%%%%%%%%%%%%%%%%%%%%%%%%%%%%%%%%%%
\section{Speculations on model realisations of MED/EDM}
%%%%%%%%%%%%%%%%%%%%%%%%%%%%%%%%%%%%%%%%%%%%%%%%%%%%%%%
\label{sec:models}

Before concluding I give here some speculative thoughts on possible
reasons for a power law scaling of sterile neutrino masses or
mixing. Without doing any detailed model building I just mention a few
possibilities where such a behaviour might occur. 

Indeed, energy dependent neutrino masses and mixing are a very
familiar phenomenon in the framework of the standard matter
effect~\cite{MSW}. Since the effective matter potential depends on the
neutrino energy the mass eigenstates and mixing angles in matter
depend on the energy. An analogous mechanism would be at work for the
sterile neutrino if it interacts with some un-known background field
or has some special interactions with standard matter.  Such a
possibility has been noted in Ref.~\cite{Strumia:2006db} and explored
recently in Ref.~\cite{Nelson:2007yq}. The interaction postulated for
the sterile neutrino should be several orders of magnitude stronger
than usual weak interactions in order to be relevant at the short
baselines in LSND or MiniBooNE. In these models neutrinos and
anti-neutrinos interact differently and therefore they will have a
different energy dependence. This may change the null-prediction for
the MiniBooNE search with anti-neutrinos mentioned in the previous
section.

An energy dependence similar to a matter potential occurs also in the
model of Ref.~\cite{Pas:2005rb}, where sterile neutrinos are allowed
to take shortcuts through particularly shaped extra
dimensions. Effectively this leads to a modification of the dispersion
relation of the sterile neutrinos which introduces a non-standard
energy dependence on active--sterile oscillations. In general also a
violation of the Lorentz symmetry such as considered for example in
Ref.~\cite{lorentz} leads to oscillations with an energy dependence
different from the standard one.

Another motivation for energy dependent neutrino masses and/or mixing
could be the idea of ``unparticle'' physics~\cite{Georgi:2007ek}. One
assumes the existence of a scale invariant sector with a non-trivial
infrared fixed point coupled to the Standard Model through
non-renormalizable operators. Such operators may have large anomalous
dimensions and hence introduce power-law running of coupling
constants. 
For example, suppose a fermionic unparticle operator $\ON$ with
mass-dimension $d_N$, with $3/2 < d_N < 5/2$, which has the quantum
numbers of a right-handed neutrino (and hence is a gauge singlet).
Then one can write a ``mass term'' $m_\alpha \overline{\nu}_\alpha
\ON$, where $\nu_\alpha$ can be either an active left-handed neutrino,
or a ``standard'' sterile neutrino.  The dimension of $m_\alpha$ is
$5/2 - d_N$ with the anomalous dimension $3/2 - d_N$.
Assuming that the coupling of $\ON$ with $\nu_\alpha$ is scale
invariant implies that the effective ``masses'' at two energy scales
$\Lambda_1$ and $\Lambda_2$ are related by $m_\alpha(\Lambda_1) =
(\Lambda_2/\Lambda_1)^{d_N-3/2} m_\alpha(\Lambda_2)$. This resembles
the scaling of Eq.~(\ref{eq:scaling}), relating the phenomenological
parameter $r$ in the above analysis to the anomalous dimension of the
unparticle operator. The physical neutrino masses have to be found as
poles in the corresponding propagator. These arguments provide a hint
that the unparticle framework might lead to the exoting energy
dependence of Eq.~(\ref{eq:scaling}); whether it is indeed possible to
obtain a valid model for MED and/or EDM active--sterile neutrino
oscillations using unparticles needs further investigation, which is
beyond the scope of this work.

Via the so-called AdS/CFT correspondence effects from a conformal
sector such as mentioned above might actually have an interpretation
also in theories with extra spacetime dimensions. In such models
couplings can exhibit power law running~\cite{Dienes:1998vh}. If the
neutrino mass is generated through a mechanism involving extra
dimensions~\cite{Dienes:1998sb} their masses and mixing may depend on
energy through these running effects. Usually the scale of new physics
in extra dimensional models is around or above the TeV energy
scale. In order to be relevant for the LSND/MiniBooNE problem one has
to assume that the mechanism responsible for the power law running can
be extended to the energy scale relevant for the experiments under
consideration (MeV to GeV) in the active--sterile neutrino sector.

At this point I will not go into further details and leave the
question whether indeed a full model for MED or EDM oscillations can
be constructed from any of the mentioned mechanisms open for future
work. I add that in a given realisation the energy dependence might be
different than assumed in Eq.~(\ref{eq:scaling}). However, the generic
assumption of a power law should be a reasonable approximation in many
cases and capture the relevant phenomenology.

%%%%%%%%%%%%%%%%%%%%%%%%%%%%%%%%%%%%%%%%%
\section{Conclusions}
%%%%%%%%%%%%%%%%%%%%%%%%%%%%%%%%%%%%%%%%%
\label{sec:summary}

I have considered short-baseline (SBL) neutrino oscillation data
including LSND and MiniBooNE in the framework of sterile neutrino
oscillations, assuming that the properties of the sterile neutrino
depend on its energy in a rather exotic way. Along these lines I
considered two different scenarios. First, I have assumed that the
mass of the sterile neutrino scales with its energy as $1/E_\nu^r$
($0\le r \le 1$). This introduces a modified energy dependence (MED)
in oscillations: Instead of the standard $1/E_\nu$ dependence one
obtains a MED with $1/E_\nu^{1+r}$. Second, I have assumed an energy
dependent mixing (EDM) of the sterile neutrino with the active ones,
namely that the elements of the mixing matrix $|U_{e4}|^2$ and
$|U_{\mu4}|^2$ scale with $1/E_\nu^r$.
In a given model realisation one can expect that both, masses as well
as mixing depend on energy in a correlated way. Here I have not
specified any underlying theory, and a phenomenological analysis has
been performed assuming the presence of either MED or EDM separately,
to show the impact on the global fit of SBL data. For a given model it
is easy to generalise the analysis and estimate the effect of the
simultaneous scaling of masses and mixing.

I find that under the hypothesis of MED or EDM oscillations LSND and
MiniBooNE data become fully consistent, and the bound from
disappearance data overlaps with the LSND allowed region.  The global
fit including all relevant appearance and disappearance experiments
improves by 12.7 (MED) or 14.1 (EDM) units in $\chi^2$ with respect to
the standard (3+1) oscillation case, and the consistency of LSND with
no-evidence appearance experiments and with disappearance experiments
improves from $4\times 10^{-5}$ for standard oscillations to 3\% (MED)
or 5\% (EDM). If the tension between LSND and KARMEN is removed from
the analysis perfect consistency of all data is found, with
probablities of 18\% and 35\% for MED and EDM, respectively, whereas
the probablitiy of standard oscillations remains unacceptably low. 
Consistency of the global data is obtained in the MED
framework by shifting the sensitivity of high energy experiments like
MiniBooNE and CDHS to larger values of $L/E_\nu$ with respect to low
energy experiments like LSND and Bugey.  In the case of EDM
oscillations the sensitivity of high energy experiments to the mixing
angle gets weaker compared to low energy experiments, leading to
consistency of all data.

In summary, I have shown that under the assumption of a non-standard
energy dependence of sterile neutrino oscillations the description of
global SBL data is significantly improved. This result is based on the
fact that various experiments operate at different energy regimes.

\bigskip

{\bf Acknowledgment.}
I would like to thank Sacha Davidson for initiating this analysis, and
for lots of discussions. Furthermore, I thank Giacomo Cacciapaglia,
Roberto Contino, Alan Cornell, Naveen Gaur, J\"orn Kersten, and
Riccardo Rattazzi for discussions on the possibility to obtain MED
and/or EDM like in Eq.~(\ref{eq:scaling}) using the concept of
un-particle physics.

%%%%%%%%%%%%%%%%%%%%
%%  BIBLIOGRAPHY  %%
%%%%%%%%%%%%%%%%%%%%

\end{document}